# Correlating Neutrino Flux and Mental Activity in the Super-Kamiokande-I Experiment

Lasse E. Bergman
University of California, San Francisco ( retired )
San Francisco, California 94143

7 August 2017

# Abstract

A search for a non-random co-variation between the Neutrino Flux and Mental Activity was undertaken, using the 5-day period version of the SK-I data taken from May 31$^{st}$, 1996 to July 15$^{th}$, 2001.

For the whole 1996-2001 period, a significant correlation between the Neutrino Flux and Mental Activity was found for each of the two midnight hours, i.e. for each of the two hours between 23:00 - 1:00 .

A significant correlation was also found across all hours for the whole year of 1998. It is not clear why this particular year should have been different from the other calendar years of 1996-2001. More specifically, significant correlations were also found for each of its hours around noon, i.e. for each of the hours between 10:00 - 16:00, for the year of 1998.

An attempt was made to interpret why significant correlations were found only for midnight hours and hours around noon, but not for any other hours: when the sun-earth axis and the staff-detector axis coincided, the likelihood increased that staff neutrinos (if any) were counted - as solar neutrinos. Unruh's second quantization in the Kerr metric was suggested as an example of theoretical support for this conjecture that Mental Activity is a quantum-gravitational process which influences the Neutrino Flux.

# I. INTRODUCTION

This paper is almost identical to a previously archived 2005 paper [1]. Its main difference from [1] is that the key expression "Staff Presence" in [1] has been replaced by the more narrow expression "Mental Activity" in this paper. Thus by definition, "Mental Activity" is a narrow subset of "Staff Presence". And by conjecture, "Mental Activity" is a quantum-gravitational process that influences the Neutrino Flux.

Among its many accomplishments, the Super-Kamiokande Collaboration [2,3,4] has placed precious neutrino data sets in the public domain [5]. As described below, it was with some trepidation that the author used one of those data sets in order to try to disclose a non-random co-variation between the Neutrino Flux and Mental Activity, with the sincere hope of not breaching that public trust.

The Collaboration has searched for but not found any significant periodicity in the SK-I neutrino flux [5], thus ruling out "semiannual (seasonal) variations of the observed solar neutrino flux because of the changing magnetic field caused by the 7.25 degree inclination of solar axis with respect to the ecliptic plane" and any "short-time variation … due to the 27-day rotation of the Sun". So, with the exception of the Homestake experiment [6], "Kamiokande and other experiments have not provided any evidence for a time variation of the neutrino flux outside of statistical fluctuations". [5]

# II. SK-I 5-DAY LONG SAMPLED SOLAR NEUTRINO DATA

The particular data set used in this paper was "collected at SK from May 31st, 1996 to July 15th, 2001, yielding a total detector live time of 1,496 days. This data taking period is known as SK-I", yielding some 15 events per day. The data set is the one arranged as 5-day periods, i.e. "neutrino data, acquired over 1,871 elapsed days from the beginning of data-taking, … divided into roughly … [5-day] long samples as listed in Table … " ( TABLE II in [5] ). The neutrino flux values have less statistical uncertainty in the 10-day long periods in TABLE I in [5] compared to the 5-day long periods in TABLE II in [5]. However the 5-day long periods were chosen for this paper because they would constitute a much larger (358 vs. 184 for the whole 1996-2001 period) number of sampling units of observation (observing the Neutrino Flux and Mental Activity in each sampling unit) and therefore would drastically reduce the estimated standard error of the sampling distribution of correlation coefficients. Below, the word "sample" refers to a collection of 5-day long periods, i.e. a collection of observation units, while the phrase "5-day long period" refers to the Collaboration's "5-day long sample".

With the choice of simple linear regression [8] as the co-variation model in this paper, the research hypothesis (the Alternative Hypothesis, in statistical terms) predicts that there exists a correlation between the Neutrino Flux and Mental Activity. The Null Hypothesis (in statistical terms), which might or might not be rejected due to the value and sample size of an empirical correlation coefficient, postulates that the true correlation is zero in the theoretical population of observation units (population of 5-day long periods) from which the sample was (postulated to be) randomly drawn. The goal in this paper is to test the null hypothesis that the Neutrino Flux and Mental Activity are linearly independent, to find out what sample correlations (if any) are so significantly high that they were hardly produced by chance alone. Of course if one would randomly draw many (say 100) samples and calculate the correlation coefficient for each of them, the result might be a collection of values where some (say 5) would be significantly high due to chance alone, i.e. even if the true correlation in the population is zero. However in TABLE I in this paper the risk is minor for overemphasizing such high-valued but phony correlations: rather than being spread arbitrarily across the table, all the significant correlations are clustered neatly around midnight hours or noon hours, and they are all positive rather than half of them being negative.

At first thought one might prefer, for correlation reasons, to be permitted access to un-binned data, e.g. to the date and time for each individual neutrino event. On second thought, however, there are good reasons why no event-by-event summary of the SK-I data is yet publicly available. As the author was graciously informed, the rationale for not releasing un-binned SK-I data publicly might roughly be expressed follows: "At the event level, interpretation of systematic errors, calibrations, and background subtraction, gets quite complicated, such that for someone not close to the gory details of the detector and reconstruction software, doing a proper job gets very difficult. This seems to be a common problem in high energy physics, unlike astronomy where raw data is routinely made public." [9] .

## III. DEFINING A CRUDE "MENTAL ACTIVITY" MEASURE

Despite the reported [5] lack of time variation in the SK-I Neutrino Flux itself, the goal of this paper is to disclose an existing correlation (if any), between the Neutrino Flux and a Mental Activity measure. As a crude definition, a particular hour-type of "Mental Activity" measure (one hour-type out of the 1+24 types 0:00-23:59;

0:00-0:59, 1:00-1:59,  . . . ,  23:00-Midnight ) for a 5-day period is specified as the  5-day period's particular type of Mental Activity minutes    (e.g. hour-type 11:00-noon minutes, in TABLE I) divided by the 5-day period's total Calendar Time minutes.   Mental Activity minutes always refer to Mon-Fri weekdays, i.e. to only "Monday through Friday, without any exceptions" .

For instance, one typical 5-day long period is period# 66 in TABLE II in [5].  One "Mental Activity" measure (hour-type 0:00-0:59) in this period is one of the 358 Mental Activity measures on which the significant (2%) correlation  0.128 ( hour-type 0:00-0:59 for 1996-2001) in TABLE I is based. This particular "Mental Activity" measure for period# 66 was calculated as follows. Its "Mental Activity" time is  209 minutes =  29 ( 0:31-0:59 Wed 4/30/97) + 60 ( 0:00-0:59 Thu 5/1/97) + 60 ( 0:00-0:59 Fri 5/2/97) + 0 (Sat 5/3/97) + 0 (Sun 5/4/97)  + 60 ( 0:00-0:59 Mon 5/5/97) .  Its corresponding Calendar Time  is 8,106 minutes = 1,409 ( 0:31-23:59 Wed 4/30/97) + 1,440 ( 0:00-23:59 Thu 5/1/97) + 1,440 ( 0:00-23:59 Fri 5/2/97) + 1,440 (0:00-23:59 Sat 5/3/97) + 1,440 (0:00-23:59 Sun 5/4/97)  + 937 ( 0:00-15:37 Mon 5/5/97).  The resulting "Mental Activity" measure (type 0:00-0:59) for period# 66 is 0.0258 = 209 minutes / 8,106 minutes.  The Neutrino Flux used for this period# 66 (and for all 1+24 hour-types in period# 66) is 2.59 (not 2.59 +0.50 -0.43, thus ignoring the statistical uncertainty until a better future definition of  "Mental Activity" would be agreed upon).

For further illustration, another typical 5-day long period is period# 124  in TABLE II in [5].  One "Mental Activity" measure (hour-type noon-12:59) in this period is one of the 72 Mental Activity measures on which the significant (1%) correlation  0.350 ( hour-type noon-12:59 for 1998) in TABLE I is based. This particular "Mental Activity" measure for period# 124 was calculated as follows. Its "Mental Activity" time is  170 minutes =  50 ( 12:10-12:59 Thu 3/5/98) + 60 ( noon-12:59 Fri 3/6/98) + 0 (Sat 3/7/98) + 0 (Sun 3/8/98) + 60 (noon-12:59 Mon 3/9/98)  + 0 ( 0:00-6:37 Tue 3/10/98) . The corresponding Calendar Time is  6,867 minutes =  710 ( 12:10-23:59 Thu 3/5/98) + 1,440 ( 0:00-23:59 Fri 3/6/98) + 1,440 ( 0:00-23:59 Sat 3/7/98) + 1,440 (0:00-23:59 Sun 3/8/98) + 1,440 (0:00-23:59 Mon 3/9/98)  + 397 ( 0:00-6:37 Tue 3/10/98) . The resulting  "Mental Activity" measure (type noon-12:59) for period# 124 is 0.0248 = 170 minutes / 6,867 minutes.  The Neutrino Flux used for this period# 124 (and for all 1+24 hour-types in period# 124)  is 2.39 (not 2.39 +0.48 -0.42, thus ignoring the statistical uncertainty for the reason mentioned above).

## IV. RESULTS

In the columns "1996-2001" and "1998" in TABLE I, significant correlation coefficient values are underlined, together with their significance levels: 1%, 2%, or 5%. The significance level denotes the so-called Type I error [10], i.e. the risk of drawing a wrong conclusion that the correlation between Neutrino Flux and Mental Activity in a theoretical population of 5-day periods is <u>not</u> zero. In order to determine the risk of wrongly saying that the correlation coefficient (actually the correlation in the theoretical population) is truly different from zero (positive or negative), a two-tailed t-test [10] was carried out:

$$t = (r * \sqrt{(n-2)}) / \sqrt{(1 - r^2)}$$

where <u>r</u> is the sample's Pearson product moment correlation coefficient and <u>n</u> is the number of 5-day periods in the sample (358 periods in the column "1996-2001" and 72 periods in the column "1998" in TABLE I). As can been seen in TABLE I, the significant correlations cluster around midnight for the whole SK-I data set of 1996-2001, and around noon-time for the year of 1998. Also, the whole year of 1998 has a significant correlation, across all 24 hours. Any reason why this calendar year of 1998 in TABLE I should have been different from the other calendar years of 1996-2001, i.e. has significant correlations, is beyond the author's knowledge. But, one might perhaps allow oneself to speculate that 1998 was a particularly busy year, with many staff on site during this early part of the SK-I experiment.

## V. DISCUSSION

It is tempting to try to interpret why significant correlations were found only for midnight hours and hours around noon, but not for any other hours: when the sun-earth axis and the staff-detector axis coincided, the likelihood increased that staff neutrinos (if any) were counted - as solar neutrinos. One may argue that there are exceptions to the rule that "Mental Activity" (i.e. staff time) means "Monday through Friday, without any exceptions", because there are other factors that count, like holidays, the number of staff, their proximity to the detector, etc. .

One should, eventually, consider such factors in re-defining the Mental Activity measure. For correlation reasons, one might also consider including in the Neutrino Flux some (non-midnight and non-noon) neutrino events where electrons did not scatter in the direction prescribed by the sun-earth axis, but in a direction that might have indicated neutrino origins in staff locations. Such a re-definition of the Neutrino Flux, and of the Mental Activity measure, would require an intimate knowledge of the un-binned data for the individual neutrino events, and of the detailed whereabouts of the staff.

Also, one may reasonably argue that the conjecture that Mental Activity influences the Neutrino Flux is far-fetched (of course a third common co-factor could be causing a significant correlation, which then would be spurious [8]) . However, theoretical justifications abound for such a conjecture, spanning the whole range from classical to quantum-gravitational physics. They all boil down to two central themes: the focus on the Planck length of ~$10^{-33}$ CM, and the still un-answered questions in quantum gravity. The elaboration of such classical, semi-classical, and quantum-gravitational justifications for the conjecture is outside the scope of this paper, but the following example gives at least the flavor of the reasoning behind it.

For instance, Unruh's second quantization in the Kerr metric [11] lends theoretical support to neutrino emission. "For a rotating hole UNRUH has shown that spontaneous creation of particles (*i.e.* neutrinos) in the hole's exterior will spin it down in a time scale $T = 10^{-43}( M / 10^{-5} g )^3$ **s.** ." [12], where M is the hole's mass energy. "Our conclusions are very similar to those for the scalar case. There will be a net spin down of the black hole on a time period of order $1/M^3$ because of neutrino emission. ... One would expect that in a stationary metric such as the Kerr metric, quantum vacuum instabilities could arise, leading to a spin down of the black hole. Again these effects would probably be significant only for small black holes (but still appreciably larger than the Planck mass black holes, i.e., $10^{-5}$ g). The ultimate resolution may be possible only when some theory is found in which both quantum mechanics and the principles of general relativity can be united" [11] .

Finally, objections to the conjecture that Mental Activity increases the Neutrino Flux may be mitigated, by inspecting the 7*24 = 168 hourly estimates of the SK-I Neutrino Flux, for the average week of 1996-2001. In [13] such a weekly 7*24 pattern proves that the five Monday thru Friday weekdays (defined as Staff time, above in this paper) have a significantly higher Neutrino Flux than the two Saturday-Sunday weekend days. Thus, Mental Activity might be said to add something locally, beyond the Collaboration's statement that "Kamiokande and other experiments have not provided any evidence for a time variation of the neutrino flux outside of statistical fluctuations" [5] .

_______________________________________

TABLE I. Empirical Correlation Coefficients as simple linear regression measures of the co-variation between the Neutrino Flux and Mental Activity, for the SK-I solar neutrino data divided into 5-day long periods. Underlined coefficients denote significant correlations and percent values denote their level of significance. Please see text for more details.

| Hour-type of Staff Time | All SK-I yrs 1996-2001 | | 1/2 yr 1996 | Yr 1997 | Yr 1998 | | Yr 1999 | Yr 2000 | 1/2 yr 2001 |
|---|---|---|---|---|---|---|---|---|---|
| 0:00-23:59 | 0.072 | | 0.189 | 0.062 | *0.242* | 5% | 0.071 | -0.080 | 0.026 |
| 0:00- 0:59 | *0.128* | 2% | 0.082 | 0.100 | 0.158 | | 0.038 | 0.233 | 0.146 |
| 1:00- 1:59 | 0.083 | | 0.083 | 0.100 | 0.158 | | 0.066 | 0.028 | 0.083 |
| 2:00- 2:59 | 0.023 | | 0.077 | 0.047 | 0.143 | | 0.045 | -0.107 | -0.019 |
| 3:00- 3:59 | 0.011 | | 0.094 | 0.026 | 0.131 | | 0.048 | -0.113 | -0.062 |
| 4:00- 4:59 | 0.022 | | 0.206 | 0.009 | 0.141 | | 0.088 | -0.154 | -0.073 |
| 5:00- 5:59 | 0.011 | | 0.156 | 0.044 | 0.175 | | 0.035 | -0.178 | -0.083 |
| 6:00- 6:59 | 0.007 | | 0.168 | 0.033 | 0.172 | | 0.032 | -0.178 | -0.088 |
| 7:00- 7:59 | 0.019 | | 0.225 | 0.020 | 0.198 | | 0.063 | -0.180 | -0.105 |
| 8:00- 8:59 | 0.020 | | 0.275 | 0.023 | 0.163 | | 0.075 | -0.198 | -0.079 |
| 9:00- 9:59 | 0.049 | | 0.251 | 0.022 | 0.177 | | 0.094 | -0.143 | 0.006 |
| 10:00-10:59 | 0.068 | | 0.246 | 0.028 | *0.263* | 5% | 0.072 | -0.106 | 0.001 |
| 11:00- noon | 0.075 | | 0.232 | 0.047 | *0.314* | 1% | 0.062 | -0.135 | 0.020 |
| noon-12:59 | 0.069 | | 0.181 | 0.044 | *0.350* | 1% | 0.028 | -0.129 | 0.023 |
| 13:00-13:59 | 0.074 | | 0.169 | 0.043 | *0.344* | 1% | 0.046 | -0.117 | 0.027 |
| 14:00-14:59 | 0.093 | | 0.203 | -0.001 | *0.299* | 2% | 0.087 | -0.112 | 0.139 |
| 15:00-15:59 | 0.096 | | 0.212 | -0.025 | *0.249* | 5% | 0.096 | -0.095 | 0.187 |
| 16:00-16:59 | 0.078 | | 0.147 | -0.027 | 0.231 | | 0.117 | -0.087 | 0.155 |
| 17:00-17:59 | 0.033 | | 0.151 | -0.008 | 0.192 | | 0.070 | -0.140 | 0.031 |
| 18:00-18:59 | 0.070 | | 0.156 | 0.069 | 0.197 | | 0.046 | -0.002 | 0.028 |
| 19:00-19:59 | 0.072 | | 0.145 | 0.100 | 0.183 | | 0.006 | 0.038 | 0.015 |
| 20:00-20:59 | 0.080 | | 0.145 | 0.127 | 0.159 | | 0.047 | 0.033 | 0.026 |
| 21:00-21:59 | 0.092 | | 0.145 | 0.133 | 0.160 | | 0.096 | 0.035 | 0.033 |
| 22:00-22:59 | 0.102 | | 0.145 | 0.153 | 0.191 | | 0.071 | 0.042 | 0.057 |
| 23:00-23:59 | *0.121* | 5% | 0.145 | 0.171 | 0.156 | | 0.070 | 0.088 | 0.122 |
| Number of 5-day periods: | 358 | | 41 | 69 | 72 | | 70 | 69 | 37 |
| Period# TABLE II Ref.[5]: | 1-358 | | 1-41 | 42-110 | 111-182 | | 183-252 | 253-321 | 322-358 |